\documentclass[aps,prd,preprint,fleqn,showkeys,showpacs,superscriptaddress]{revtex4}
\usepackage[colorlinks=true]{hyperref}
\hypersetup{colorlinks=true, citecolor=blue, urlcolor=blue, linkcolor=blue}

\usepackage{graphicx}
\usepackage{tabulary}
\usepackage{multirow,tabularx}
\usepackage{amssymb}
\usepackage{amsmath}
\usepackage{bm}
\usepackage{enumitem}
\usepackage{etoolbox}

\usepackage{xcolor}
\makeatletter

\patchcmd\frontmatter@PACS@format{\addvspace{11\p@}}{}{}{}
\pretocmd\frontmatter@keys@format{\addvspace{11\p@}}{}{}
\patchcmd{\titleblock@produce}
  {\@pacs@produce\@pacs\@keywords@produce\@keywords}
  {\@keywords@produce\@keywords\@pacs@produce\@pacs}
  {}{}
\makeatother

\begin{document}
\title{Decay properties of beauty and charm mesons within Isgur-Wise function formalism}

\author{C. W. Xiao}
\affiliation{School of Physics and Electronics, Central South University, Changsha 410083, China}

\author{S. Rahmani}
\email{s.rahmani120@gmail.com}
\affiliation{School of Physics and Electronics, Central South University, Changsha 410083, China}

\author{H. Hassanabadi}
\affiliation{Faculty of Physics, Shahrood University of Technology, P. O. Box: 3619995161-316, Shahrood, Iran.}

\begin{abstract}
We investigate the decay properties of some beauty and charm mesons with a phenomenological potential model. First, we consider the nonrelativistic Hamiltonian of the mesonic system with Coulomb plus exponential terms and study the wave function and the energy of the system using the variational approach. Thereby, we compute the masses, the decay constants, the leptonic branching fractions of heavy-light mesons and the mixing mass parameter $\Delta {m_{{B_q}}}$. We study the radiative leptonic decay widths of ${D_s} \to \gamma \ell \bar \nu $, ${D^ - } \to \gamma \ell \bar \nu $ and the semileptonic decay widths of ${\bar B_{(s)}} \to {D_{(s)}}\ell \bar \nu $, ${\bar B_{(s)}} \to D_{(s)}^*\ell \bar \nu $. Using Isgur-Wise functions, we calculate the branching ratios of $B \to {D^{(*)}}\pi $ and two-body nonleptonic decay of $D \to K\pi $. Our results are consistent with other theoretical models and the experimental results.
\end{abstract}

\maketitle
\thispagestyle{empty}

\section{Introduction }
In recent years, the study of beauty and charm mesons have been remarkably achieved both in the experimental and theoretical sides. But, there are still some challenges and open questions. In the study of semileptonic decays of $B$ mesons which contain the flavour-changing quark transitions $b \to c$, one needs to introduce a universal Isgur-Wise function (IWF) $\xi (v.v')$. The semilptonic $B$ decays provide a good opportunity to measure the Cabibbo-Kobayashi-Maskawa (CKM) matrix element $|V_{cb}|$ and the knowledge of the heavy-light meson dynamics. Besides the study of two-body nonleptonic decays of $B$ and $D$ mesons are also caught great of interests in particle physics because they contain valuable information on the electroweak interactions of quarks, flavor mixing, $CP$ violation \cite{Fusheng:2011tw} and they are useful for testing some QCD motivated models. The semileptonic and nonleptonic decay widths with $B \to {D^{(*)}}$ are expressed in terms of IWF in HQET. There are different parameterizations for IWF within phenomenological approaches \cite{Hioki:1993yr,LeYaouanc:2003rn,Coleman:2000gu,Zhao:2006at}. 

The decay constants of heavy-light mesons play an important role in particle physics such as the ${B^0} - {\bar B^0}$ mixing, the treatment of nonleptonic heavy flavour decays in the factorization approximation, the analysis of the CKM matrix element and also the connection of the  leptonic constants to the wave function at the origin in the nonrelativistic quark model \cite{Paver:1992cb,Wang:2005qx,Yang:2011ie,Chang:2018aut}. 

Up to now, there have been some valuable studies on the weak decays of $B$ and $D$ mesons using the factorization approach \cite{Fusheng:2011tw,Hioki:1993yr}, QCD sum rules \cite{Azizi:2008tt,Azizi:2008vt}, Bethe-Salpeter equation approach \cite{Ivanov:1998ya,AbdElHady:1997rj} and non-relativistic constituent quark model \cite{Lu:2002mn}. Li et al studied both the semileptonic decays of $\bar B_s^0 \to D_s^ + \ell {\bar \nu _\ell }$  and nonleptonic decays $B_s^{} \to D_s^ + M$ where $M$ is a light or charmed meson under the factorization approach \cite{Li:2009wq}. Ivanov et al analyzed the exclusive leptonic and semileptonic $B$ decays $B \to \ell {\nu _\ell }$ and $B \to {D^{(*)}}\ell {\nu _\ell }$ with the covariant quark model \cite{Ivanov:2015tru}. The Non-leptonic decays of $B$ mesons into two mesons studied by Kramer and Lü using two versions of pole-dominance model in addition with a factorization assumption \cite{Kramer:1997yh}. With a relativistic potential model, Sun and Yang obtained the wave functions and leptonic decays of bottom mesons \cite{Sun:2019xyw}. Lü and Song calculated the branching fractions of $D_{(s)}^ -  \to \gamma \ell \bar \nu (\ell  = e,\mu )$  using the non-relativistic constituent quark model and the effective Lagrangian for the heavy ﬂavor decays \cite{Lu:2002mn}. The spectroscopy and the decay properties of $B$ and $B_s$ mesons have been studied using Hydrogenic and Gaussian wave functions \cite{Devlani:2012zz}. 

In next section, we make a brief introduction of the Hamiltonian of a mesonic system using of a variation method. In section \ref{3}, we obtain the masses of pseudoscalar and vector mesons in beauty and charm sectors, the leptonic decay widths of $B$ and $D$ mesons. Using of their decay constants, we also evaluate the radiative leptonic decay widths of charm mesons including $D$ and $D_s$. The mixing mass parameter of $B$ and $B_s$ mesons are given in the next section. Following, we use two different parameterizations of IWF and analyze semileptonic decay widths of $B$ mesons in section \ref{5}. Then, using of differential semileptonic decay widths of $B$ mesons at maximum recoil, we study the nonleptonic decays of $B$ to $D$ and a pion in their final transitions in next section. $D$ decays to a pion and a $K$ meson are also included. At the end, we present our conclusions and outlook. 

\section{General framework }
\begin{figure}
\centering
\includegraphics[width=0.48\linewidth]{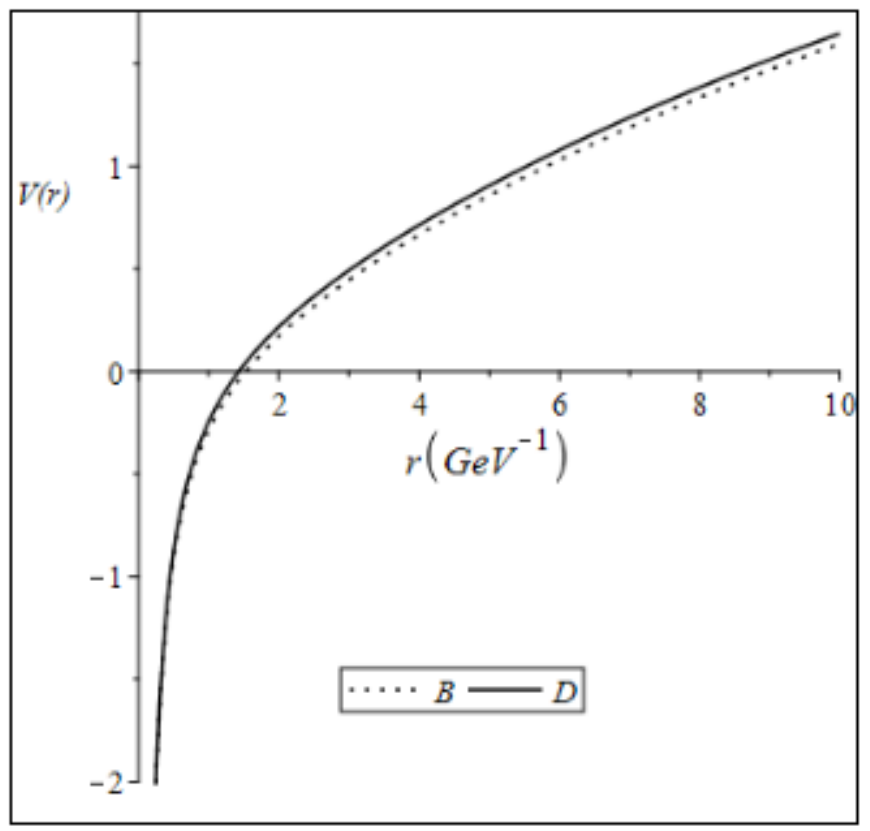}
\caption{Behavior of the potential for $B$ and $D$ mesons.}
\label{fig:vr}
\end{figure}

\begin{figure}
\centering
\includegraphics[width=0.48\linewidth]{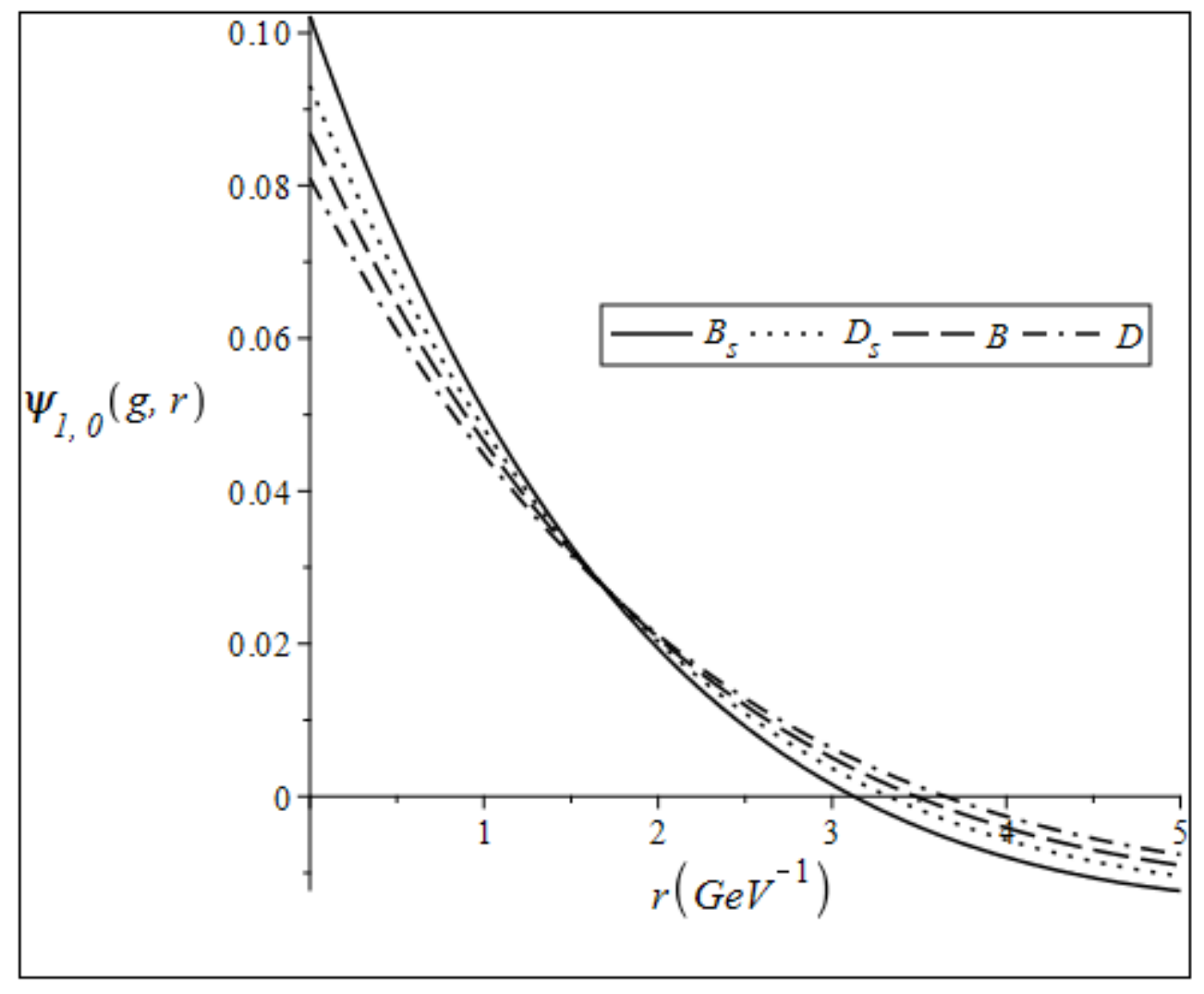}
\caption{Wave functions for beauty and charm mesons.}
\label{fig:wave}
\end{figure}

For the heavy-light bound states of $B$ and $D$ mesons, we consider the nonrelativistic Hamiltonian to obtain the wave function of the mesonic system,
\begin{equation}
    H = \frac{{{P^2}}}{{2\mu }} + V(r),
\end{equation}
where the potential is assumed as the Coulomb plus exponential terms \cite{Pang:2019ttv,Chang:2014jca}, 
\begin{equation}
    V(r) = \left[ {\frac{{ - 3}}{4}\left( {\frac{{b(1 - {e^{ - \lambda r}})}}{\lambda } + {V_0}} \right) + \frac{{{\alpha _s}(r)}}{r}} \right]{\vec F_1}.{\vec F_2},
    \label{eq:vr}
\end{equation}
and $\mu $  is the reduced mass of the mesons. The potential parameters are $b = 0.221(Ge{V^2})$  and $\lambda  = 0.{\text{0635(}}GeV)$ \cite{Pang:2019ttv}. $V_0$ is a free parameter and will be determined by fitting to the experimental masses of the heavy-light mesons. For mesons we have $\left\langle {{{\vec F}_1}.{{\vec F}_2}} \right\rangle  =  - \frac{4}{3},$ where $\vec F$ are related to the Gell-Mann matrices. The Coloumb-like term $\frac{{ - 4{\alpha _s}(r)}}{{3r}}$  originates from the one gluon exchange diagram for the short-distance contributions, and the exponential term is for confinement effects at a long distance. Fig. \ref{fig:vr} shows the variation of considered potential, Eq. \eqref{eq:vr} for $B$ and $D$ mesons.
The value of the QCD coupling constant can be obtained through \cite{Pathak:2012kp}:
\begin{equation}
    {\alpha _s}({\mu ^2}) = \frac{{4\pi }}{{\left( {11 - \frac{{2{n_f}}}{3}} \right)\ln \left( {\frac{{{{(2\mu )}^2} + M_B^2}}{{{\Lambda ^2}}}} \right)}},
\end{equation}
where the background mass is ${M_B} = 2.24 \times \sqrt b $, ${\Lambda _{QCD}} = 0.200GeV$ and ${n_f} = 3$ \cite{Chakrabarty:1990gs}. We use the hydrogen-like wave function as a trial wave function \cite{Devlani:2012zz,Hassanabadi:2016kqq}, 
\begin{equation}
    {\psi _{n,l}}(g,r) = N{g^{\frac{3}{2}}}{(gr)^l}{e^{ - gr}}L_n^{2l + 1}(2gr),
    \label{eq:psigr}
\end{equation}
where $N$ is the normalization constant, $g$ the variational parameter and  $L$ the Laguerre polynomial. We take the quantum numbers $n=1, l=0$ in the present work. Thus by considering the condition $\int\limits_0^\infty  {4\pi {r^2}\psi _{1,0}^2(g,r)dr = 1}$, $N$ will be $\frac{1}{{2\sqrt \pi  }}$. We have plotted the wave functions for $B, B_s , D_s $ and $D$ mesons in Fig. \ref{fig:wave}, which has been normalized to one. 
By minimizing the trial energy and taking the derivative with respect to $g$, setting it equal to zero as 
\begin{equation}
    \frac{\partial }{{\partial g}}\left( { < {\psi _{n,l}}(g,r)|H|{\psi _{n,l}}(g,r) > } \right) = 0,
    \label{eq:partg}
\end{equation}
and solving for $g$, we can evaluate the energy of the mesonic system. We have shown the obtained $g$ parameter for $B$ and $D$ mesons in the second column of Table \ref{tab:mass}. $g$ depends on quantum numbers $n,l$ and masses of quarks. We have considered the ground state of mesons and obtained the variation parameter for different $D$ and $B$ mesons using Eqs. \eqref{eq:psigr} and \eqref{eq:partg}.

\begin{center}
\begin{table}[ht]
\caption{Masses and decay constants of heavy-light mesons (${V_0} = 0.014GeV$  for bottom mesons and ${V_0} = 0.064GeV$ for charmed ones)}
\label{tab:mass}
\begin{tabularx}{\textwidth}{|X|X|X|X|X|X|} 
\hline    
Meson &  $g $  &  Mass (Ours)-in GeV &   Exp.-in GeV \cite{PDG:2020} &  ${f_{P/V}}$ (Ours)-in GeV &  ${f_{P/V}}$ (Others)-in MeV\\ \hline  
   ${B^ \pm }$ & 0.287 & 5.273 & 5.279 & 0.135 & $198 \pm 14$ \cite{Yang:2011ie}, $187.2_{ - 4.3}^{ + 4.0}$ \cite{Chang:2018aut} \\  \hline
${D^ \pm }$ & 0.274 & 1.877 & 1.869 & 0.208 & 205.8 \cite{PDG:2020} \\ \hline
 $D_s^ \pm $ & 0.301 & 2.004 & 1.968 & 0.238 & $233.1_{ - 5.4}^{ + 5.0}$ \cite{Chang:2018aut} \\ \hline
 $B_s^0$ & 0.320 & 5.399 & 5.366 & 0.162 & $237 \pm 17$ \cite{Yang:2011ie} \\ \hline
 ${B^*}$ & 0.287 & 5.294 & 5.324 & 0.135 & $193.1_{ - 4.6}^{ + 4.3}$ \cite{Chang:2018aut} \\ \hline
 ${D^*}$ & 0.274 & 1.934 & 2.006 & 0.205 & $226.6_{ - 10.2}^{ + 5.9}$ \cite{Chang:2018aut} \\ \hline
 $D_s^{* \pm }$ & 0.301 & 2.060 & 2.112 & 0.235 & $254.7_{ - 6.7}^{ + 6.3}$ \cite{Chang:2018aut} \\ \hline
 $B_s^*$ & 0.320 & 5.420 & 5.415 & 0.162 & $272 \pm 20$ \cite{Wang:2005qx} \\ \hline
\end{tabularx} 
\end{table}
\end{center}

\section{Masses and leptonic branching fractions } \label{3}
The pseudoscalar or vector meson mass is taken to be 
\begin{equation}
    {M_{P/V}} = {m_1} + {m_2} + {E_{1,0}}  +  < {H_{sd}} > 
\end{equation}
where $m_1$ and $m_2$ are the quark masses and $E_{1,0}$ is the energy of the mesonic system in the ground state. The spin dependent term is given by
\begin{equation}
    \left\langle {{H_{sd}}} \right\rangle  = \left\{ \begin{gathered}
  \frac{{8\pi {\alpha _s}}}{{9{m_1}{m_2}}}|{\psi _{P/V}}(0){|^2}\,\,{\text{for S = 1}} \hfill \\
   - \frac{{8\pi {\alpha _s}}}{{3{m_1}{m_2}}}|{\psi _{P/V}}(0){|^2}\,\,{\text{for S = 0}} \hfill \\ 
\end{gathered}  \right.
\end{equation}
where the wave function at the origin is
\begin{equation}
    |{\psi _{P/V}}(0){|^2} = \frac{\mu }{{2\pi {\hbar ^2}}} < \frac{{dV(r)}}{{dr}} >. 
\end{equation}
Input values of quark masses are ${m_d} = {m_u} = 0.336GeV,{m_s} = 0.465GeV,$ ${m_c} = 1.55GeV,$ and ${m_b} = 4.97GeV$ \cite{Pathak:2012kp}. We have calculated the masses of $B$ and $D$ mesons in the third column of Table \ref{tab:mass}. As one can see from Table \ref{tab:mass}, in most cases our results are in agreement with the experimental data. For instance, the differences between our obtained masses and Ref. \cite{PDG:2020} are 6 MeV, 8 MeV, 36 MeV and 33 MeV for the pseudoscalar mesons $B$, $D$, $D_s$ and $B_s$ respectively. By using of $\sigma  = \sqrt {\frac{1}{N}\sum\limits_{n = 1}^N {{{[\frac{{{M_{Ours}} - {M_{Exp}}}}{{{M_{Exp}}}}]}^2}} } $, we have computed the root mean square deviations of our obtained masses with experimental data of Ref. \cite{PDG:2020} as $\pm0.017$.
\par
The leptonic decays of $B$ and $D$ mesons contain the flavour-changing transitions. In these process, a quark and antiquark annihilate via a virtual $W$ boson. Let $P$ be any of the pseudoscalar mesons including $B$, $D$ and $D_s$. In the Standard Model (SM), the purely leptonic decay widths of these heavy-light mesons can be obtained by
\begin{equation}
    \Gamma (P \to l\nu ) = \frac{{G_F^2|{V_{{q_1}{q_2}}}{|^2}f_P^2}}{{8\pi }}m_l^2{\left( {1 - \frac{{m_l^2}}{{M_P^2}}} \right)^2}{M_P}
    \label{eq:width}
\end{equation}
where ${V_{{q_1}{q_2}}}$ is the CKM matrix element between the quarks ${q_1}{q_2}$ inside $P$. In the case of ${D_s},D,B$, it is ${V_{cs}}$, ${V_{cd}}$ and ${V_{ub}}$, respectively. In the nonrelativistic limit, the decay constants of pseudoscalar and vector mesons can be expressed through the meson wave function at the origin by Van-Royen-Weisskopf formula \cite{Rahmani:2017vbg},
\begin{equation}
    f_{P/V}^{} = \sqrt {\frac{{12}}{{{M_{P/V}}}}} |{\psi _{P/V}}(0)|
\end{equation}
In this work, we have taken Fermi coupling constant, CKM matrix elements and the masses of leptons as follows ${G_F} = 1.16637 \times {10^{ - 5}}Ge{V^{ - 2}}$, $|{V_{ub}}| = 0.00351,|{V_{cs}}| = 0.97344,|{V_{cd}}| = 0.22520$, ${m_\mu } = 0.105GeV,{m_\tau } = 1.776GeV,{m_e} = 0.510 \times {10^{ - 3}}GeV$ \cite{PDG:2020}. The fifth column of Table \ref{tab:mass} shows our obtained values for the decay constants of the beauty and charm mesons. 
The ratio of charm decay constants reported as $\frac{{{f_{{D_s}}}}}{{{f_D}}} = {\text{1}}{\text{.175}}$  and $\frac{{{f_{{B_s}}}}}{{{f_B}}} = {\text{1}}{\text{.209}}$ for the bottom pseudoscalar-meson decay constants \cite{PDG:2020}. We have obtained $\frac{{{f_{{D_s}}}}}{{{f_D}}} = {\text{1}}{\text{.146}}$  and  $\frac{{{f_{{B_s}}}}}{{{f_B}}} = {\text{1}}{\text{.201}}$, where the differences of our mentioned values are about 2.45 $\%$ and 0.65 $\%$, respectively, compared to Ref. \cite{PDG:2020}. The results of Ref. \cite{Ebert:2006hj} are 1.15 for these fractions. Qin Chang et. al. reported $\frac{{{f_{{D_s}}}}}{{{f_D}}} = {\text{1}}{\text{.129}}$, $\frac{{{f_{{B_s}}}}}{{{f_B}}} = {\text{1}}{\text{.166}}$, $\frac{{{f_{D_s^*}}}}{{{f_{{D^*}}}}} = {\text{1}}{\text{.12}}$, $\frac{{{f_{D_{}^*}}}}{{{f_D}}} = {\text{1}}{\text{.097}}$, $\frac{{{f_{D_s^*}}}}{{{f_{{D_s}}}}} = {\text{1}}{\text{.093}}$, $\frac{{{f_{B_s^*}}}}{{{f_{{B^*}}}}} = {\text{1}}{\text{.21}}$, $\frac{{{f_{B_{}^*}}}}{{{f_B}}} = {\text{1}}{\text{.027}}$ and $\frac{{{f_{B_s^*}}}}{{{f_{{B_s}}}}} = {\text{1}}{\text{.028}}$ \cite{Chang:2018aut}. Our results for the ratios of decay constants, $\frac{{{f_{D_s^*}}}}{{{f_{{D^*}}}}} = {\text{1}}{\text{.148}}$, $\frac{{{f_{D_{}^*}}}}{{{f_D}}} = 0.{\text{985}}$, $\frac{{{f_{D_s^*}}}}{{{f_{{D_s}}}}} = 0.{\text{986}}$, $\frac{{{f_{B_s^*}}}}{{{f_{{B^*}}}}} = {\text{1}}{\text{.201}}$, $\frac{{{f_{B_{}^*}}}}{{{f_B}}} = 0.{\text{998}}$ and $\frac{{{f_{B_s^*}}}}{{{f_{{B_s}}}}} = 0.{\text{998}}$, are generally in compatible with them, where the difference of our obtained value for $\frac{{{f_{B_{}^*}}}}{{{f_B}}}$ is about 4.18 $\%$, and $\frac{{{f_{B_s^*}}}}{{{f_{{B_s}}}}}$ about 2.47 $\%$ in comparison with Ref. \cite{Chang:2018aut}. The ratios of the decay constants for the bottom mesons, $\frac{{{f_{B_{}^*}}}}{{{f_B}}} = 0.{\text{958}} \pm {\text{0}}{\text{.022}}$  and  $\frac{{{f_{B_s^*}}}}{{{f_{{B_s}}}}} = 0.{\text{974}} \pm {\text{0}}{\text{.010}}$, have also been given with LQCD approach calculations \cite{Lubicz:2016bbi}, which are similar to what we get. 
With the leptonic decay widths that we obtain from Eq. \eqref{eq:width}, the leptonic decay branching ratios for the beauty and charm mesons are given in the second column of Table \ref{tab:leptonic}, where we compare with the experimental ones \cite{PDG:2020} and find that they are consistent.

\begin{center}

\begin{table}[ht]
\caption{Leptonic decays of heavy-light mesons}
\label{tab:leptonic}
\begin{tabularx}{\textwidth}{|X|X|X|} 
\hline    

  Decay & $Br$ (Our) &  Exp. \cite{PDG:2020} \\ \hline  
   
  ${B^ + } \to {\tau ^ + }{\nu _\tau }$  & ${\text{0}}{\text{.40}} \times {10^{ - 4}}$ & $(1.09 \pm 0.24) \times {10^{ - 4}}$ \\ \hline
  
  ${B^ + } \to {\mu ^ + }{\nu _\mu }$  & ${\text{1}}{\text{.76}} \times {10^{ - 7}}$  & ${\text{2}}{\text{.90}} \times {10^{ - 7}}$ \\ \hline
  
  $ {B^ + } \to {e^ + }{\nu _e} $ & ${\text{4}}{\text{.16}} \times {10^{ - 12}}$ & $<{\text{9}}{\text{.8}} \times {10^{ - 7}}$ \\ \hline
 
 ${D^ + } \to {\tau ^ + }{\nu _\tau }$ & ${\text{2}}{\text{.28}} \times {10^{ - 2}}$ & $(1.20 \pm 0.27) \times {10^{ - 3}}$ \\ \hline
 
 ${D^ + } \to {\mu ^ + }{\nu _\mu }$ & ${\text{7}}{\text{.22}} \times {10^{ - 3}}$ &  $(3.74 \pm 0.17) \times {10^{ - 4}}$ \\ \hline
 
 ${D^ + } \to {e^ + }{\nu _e}$  & ${\text{1}}{\text{.71}} \times {10^{ - 7}}$  & $< 8.8 \times {10^{-6}}$  \\ \hline
 
 $D_s^ +  \to {\tau ^ + }{\nu _\tau }$ & ${\text{6}}{\text{.52}}\% $
& $(5.48 \pm 0.23)\% $  \\ \hline

$D_s^ +  \to {\mu ^ + }{\nu _\mu }$ & ${\text{4}}{\text{.91}} \times {10^{ - 3}}$ & $(5.49 \pm 0.16) \times {10^{ - 3}}$  \\ \hline

$D_s^ +  \to {e^ + }{\nu _e}$ & ${\text{1}}{\text{.16}} \times {10^{ - 7}}$ & $< 8.3 \times {10^{-5}}$ \\ \hline

\end{tabularx} 
\end{table}
\end{center}
\par
Recently, Fleischer et. al. have shown their results for the branching ratios of leptonic $D_s$ decays with the SM as $Br(D_s^ +  \to {e^ + }{\nu _e}) = (1.24 \pm 0.02) \times {10^{ - 7}},Br(D_s^ +  \to {\mu ^ + }{\nu _\mu }) = (5.28 \pm 0.08) \times {10^{ - 3}}$, $Br(D_s^ +  \to {\tau ^ + }{\nu _\tau }) = (5.15 \pm 0.08) \times {10^{ - 2}}$ \cite{Fleischer:2019wlx}. Comparing with their results, there are 6.14 $\%$ differences for the leptonic decay width of $D_s^ +  \to {e^ + }{\nu _e}$,  7.07 $\%$ differences for the one of the decay $D_s^ +  \to {\mu ^ + }{\nu _\mu }$,  and 26.56 $\%$ for $D_s^ +  \to {\tau ^ + }{\nu _\tau }$. In the case of leptonic $B$-decay branching fractions $Br({B^ - } \to {e^ - }{\nu _e}) = 1.16 \times {10^{ - 11}},Br({B^ - } \to {\mu ^ - }{\nu _\mu }) = 0.49 \times {10^{ - 6}}$  and $Br({B^ - } \to {\tau ^ - }{\nu _\tau }) = 1.10 \times {10^{ - 4}}$  have been reported by Ivanov et. al. \cite{Ivanov:2015tru}. Our results in Table \ref{tab:leptonic} are consistent with them.
The deviations of our leptonic $B$-decay results in comparison with them are $\pm0.641$, $\pm0.640$ and $\pm0.639$, respectively, for $e$, $\mu $ and $\tau $.

Furtheremore,  another one of the quantities that we can calculate by the obtained masses and decay constants is the radiative leptonic decays of charm mesons \cite{Lu:2002mn}
\begin{equation}
    \Gamma ({D_s} \to \gamma \ell \bar \nu ) = \frac{{\alpha G_F^2|{V_{cs}}{|^2}}}{{2592{\pi ^2}}}f_{{D_s}}^2M_{{D_s}}^3[{x_s} + {x_c}],
\end{equation}
with
\begin{equation}
    {x_s} = {\left( {3 - \frac{{{M_{{D_s}}}}}{{{m_s}}}} \right)^2},{x_c} = {\left( {3 - 2\frac{{{M_{{D_s}}}}}{{{m_c}}}} \right)^2}.
\end{equation}
For the case of $D$ mesons, we have \cite{Lu:2002mn}
\begin{equation}
    \Gamma ({D^ - } \to \gamma \ell \bar \nu ) = \frac{{\alpha G_F^2|{V_{cd}}{|^2}}}{{2592{\pi ^2}}}f_D^2M_D^3[{x_d} + {x_c}],
\end{equation}
where
\begin{equation}
    {x_d} = {\left( {3 - \frac{{{M_D}}}{{{m_d}}}} \right)^2},{x_c} = {\left( {3 - 2\frac{{{M_D}}}{{{m_c}}}} \right)^2}.
\end{equation}
Using $\alpha  = \frac{1}{{137}}$, ${\tau _{{D_s}}} = 5.04 \times {10^{ - 13}}s,{\tau _D} = 1.04 \times {10^{ - 12}}s$ \cite{PDG:2020}, we obtain the branching ratios of radiative leptonic decays for charm mesons in Table \ref{tab:radilep}, where our results are close to the ones obtained Ref. \cite{Lu:2002mn}.
\begin{center}

\begin{table}[ht]
\caption{Branching ratios of radiative leptonic decays}
\label{tab:radilep}
\begin{tabularx}{\textwidth}{|X|X|X|} 
\hline    

  Decay & $Br$ (Our) & Results of Ref. \cite{Lu:2002mn} \\  \hline  
   
   ${D_s} \to \gamma \ell \nu $ & ${\text{2}}{\text{.43}} \times {\text{1}}{{\text{0}}^{ - 5}}$ & $1.8 \times {10^{ - 5}}$   \\ \hline  
   
    $D \to \gamma \ell \nu $ & ${\text{6}}{\text{.25}} \times {10^{ - 6}}$ & $4.6 \times {10^{ - 6}}$ \\  \hline 
 
\end{tabularx} 
\end{table}
\end{center}
\par
In PDG \cite{PDG:2020}, the branching ratio of $D \to \gamma {e^ + }{\nu _e}$ is $ < 3.0 \times {10^{ - 5}}$ . Our result of ${\text{6}}{\text{.25}} \times {10^{ - 6}}$ is in agreement with it. Also we have obtained ${\text{2}}{\text{.43}} \times {\text{1}}{{\text{0}}^{ - 5}}$ for the ratio of $D_s$ radiative leptonic decay, which is in agreement with the one obtained in \cite{PDG:2020} $Br({D_s} \to \gamma {e^ + }{\nu _e}) < 1.3 \times {10^{ - 4}}$. Furthermore, we have obtained $\Gamma (D \to \gamma \ell \nu ) = {\text{3}}{\text{.954}} \times {10^{ - 18}}GeV$  and $\Gamma ({D_s} \to \gamma \ell \nu ) = {\text{3}}{\text{.178}} \times {10^{ - 17}}GeV$,  which are in agreement with the results of Ref. \cite{Lu:2002mn}: $\Gamma (D \to \gamma \ell \nu ) = {\text{2}}{\text{.9}} \times {10^{ - 18}}GeV$ and $\Gamma ({D_s} \to \gamma \ell \nu ) = {\text{2}}{\text{.3}} \times {10^{ - 17}}GeV$. The deviations of our obtained values for branching ratios of $D \to \gamma \ell\nu $  and $D_s \to \gamma \ell\nu $ comparing with Ref. \cite{Lu:2002mn} are about $\pm0.359$ and $\pm0.353.$, respectively. 

\section{Mixing mass parameter}
Particle-antiparticle mixing phenomena have fundamental importance in testing the SM. With the calculated meson masses and pseudoscalar decay constants, we can compute the oscillation frequency $\Delta {m_{{B_q}}},q = d,s$. This parameter is a measure of the frequency of the change from $B$ into $\bar B$ and can be parameterized as \cite{Buras:2003td,Hulsbergen:2013lma,Pathak:2011km},  
\begin{equation}
    \Delta {m_B} = \frac{{G_F^2m_t^2{M_{{B_q}}}f_{{B_q}}^2}}{{8\pi }}\left[\frac{1}{4} + \frac{9}{{4(1 - {x_t})}} - \frac{3}{{2{{(1 - {x_t})}^2}}} - \frac{{3x_t^2}}{{2{{(1 - {x_t})}^3}}}\right] \eta _t|V_{tq}^*{V_{tb}}{|^2}B;\,q = d,s
\end{equation}
which depends on parameters of SM, such as $G_F$, the CKM matrix elements (${V_{td}} = 7.4 \times {10^{ - 3}},{V_{tb}} = 1,{V_{ts}} = 40.6 \times {10^{ - 3}}$), top quark (${m_t} = 174GeV$) and $W$ boson (${m_W} = 80.403GeV$) masses. Moreover, we have used ${x_t} = \frac{{m_t^2}}{{m_W^2}}$, $B = 1.34,$ representing the correction to the vacuum insertion, and the numerical factor ${\eta _t} = 0.55$ standing for QCD corrections. We show our calculated values for the mixing parameters for $B$ mesons in Table \ref{tab:mix}, where our results are found to be comparable with the other theoretical values and experimental data in Refs. \cite{PDG:2020,Hulsbergen:2013lma,Pathak:2011km}. The deviations of our results are about $\pm0.534$ and $\pm0.643$ for $B_s$ and $B_d$, respectively, compared to Refs. \cite{Hulsbergen:2013lma,PDG:2020}. 

\begin{center}

\begin{table}[ht]
\caption{Mixing parameters for $B$ mesons}
\label{tab:mix}
\begin{tabularx}{\textwidth}{|X|X|X|} 
\hline    

  Meson & $\Delta {m_B}(p{s^{ - 1}})$ (Our) & Other results  \\  \hline  
   
   ${B_s} $ & 8.06 & 23.88 \cite{Pathak:2011km}, $17.3 \pm 2.6$ \cite{Hulsbergen:2013lma}  \\ \hline  
   
    $B_d$ & 0.18 & 0.55 \cite{Pathak:2011km}, 0.507 \cite{PDG:2020}\\  \hline 
 
\end{tabularx} 
\end{table}
\end{center}

\section{Semileptonic decays of $B$ mesons} \label{5}
\begin{figure}
\centering
\includegraphics[width=0.48\linewidth]{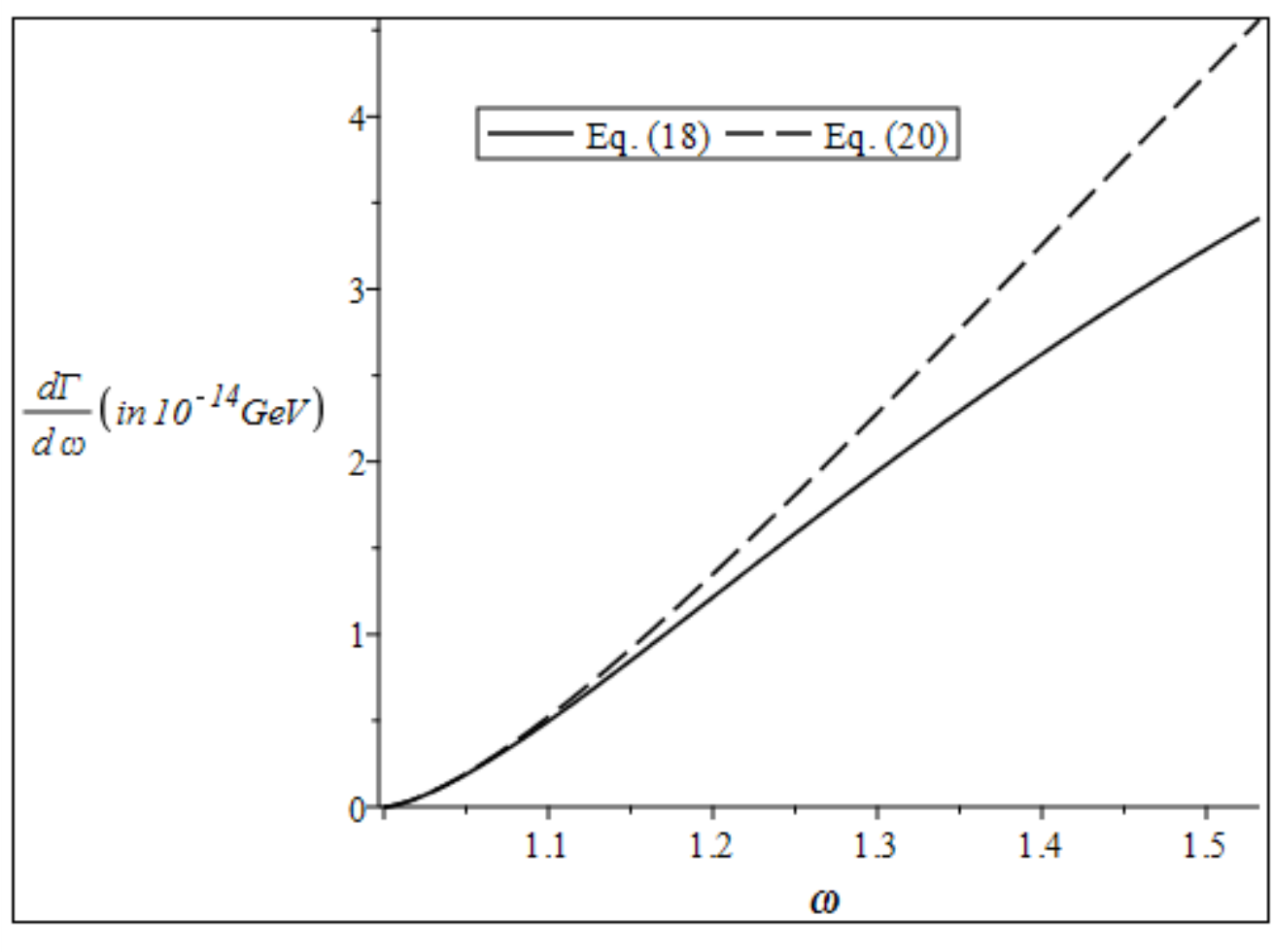}
\caption{Differential semileptonic decay widths of ${B_s} \to {D_s}\ell \nu $ versus $\omega $ for two cases of IWFs.}
\label{fig:diff}
\end{figure}

The differential semileptonic decays can be expressed \cite{Rahmani:2017exi,Hassanabadi:2014isa}
\begin{equation}
    \begin{gathered}
  \frac{{d\Gamma ({{\bar B}_{(s)}} \to {D_{(s)}}\ell \bar \nu )}}{{d\omega }} = \frac{{G_F^2|{V_{cb}}{|^2}}}{{48{\pi ^3}}}{({M_{{B_{(s)}}}} + {M_{{D_{(s)}}}})^2}M_{{D_{(s)}}}^3{({\omega ^2} - 1)^{\frac{3}{2}}} \hfill \\
   \times |{h^ + }(\omega ) - \frac{{{M_{{B_{(s)}}}} - {M_{{D_{(s)}}}}}}{{{M_{{B_{(s)}}}} + {M_{{D_{(s)}}}}}}{h^ - }(\omega ){|^2} \hfill  
   \label{eq:BtoD}
\end{gathered} 
\end{equation}
in terms of form factors ${h^ - }(\omega )$ and ${h^ + }(\omega )$, which are given by
\begin{equation}
    \begin{gathered}
  {h^ - }(\omega ) = 0, \hfill \\
  {h^ + }(\omega ) = \xi (\omega ), \hfill \\ 
\end{gathered} 
\end{equation}
in the heavy quark symmetry. In the range of semileptonic decay width, we have shown the behavior of $\frac{{d\Gamma }}{{d\omega }}({B_s} \to {D_s}\ell \nu )$ as a function of $\omega $ in Fig. \ref{fig:diff}. 

\begin{figure}
\centering
\includegraphics[width=0.48\linewidth]{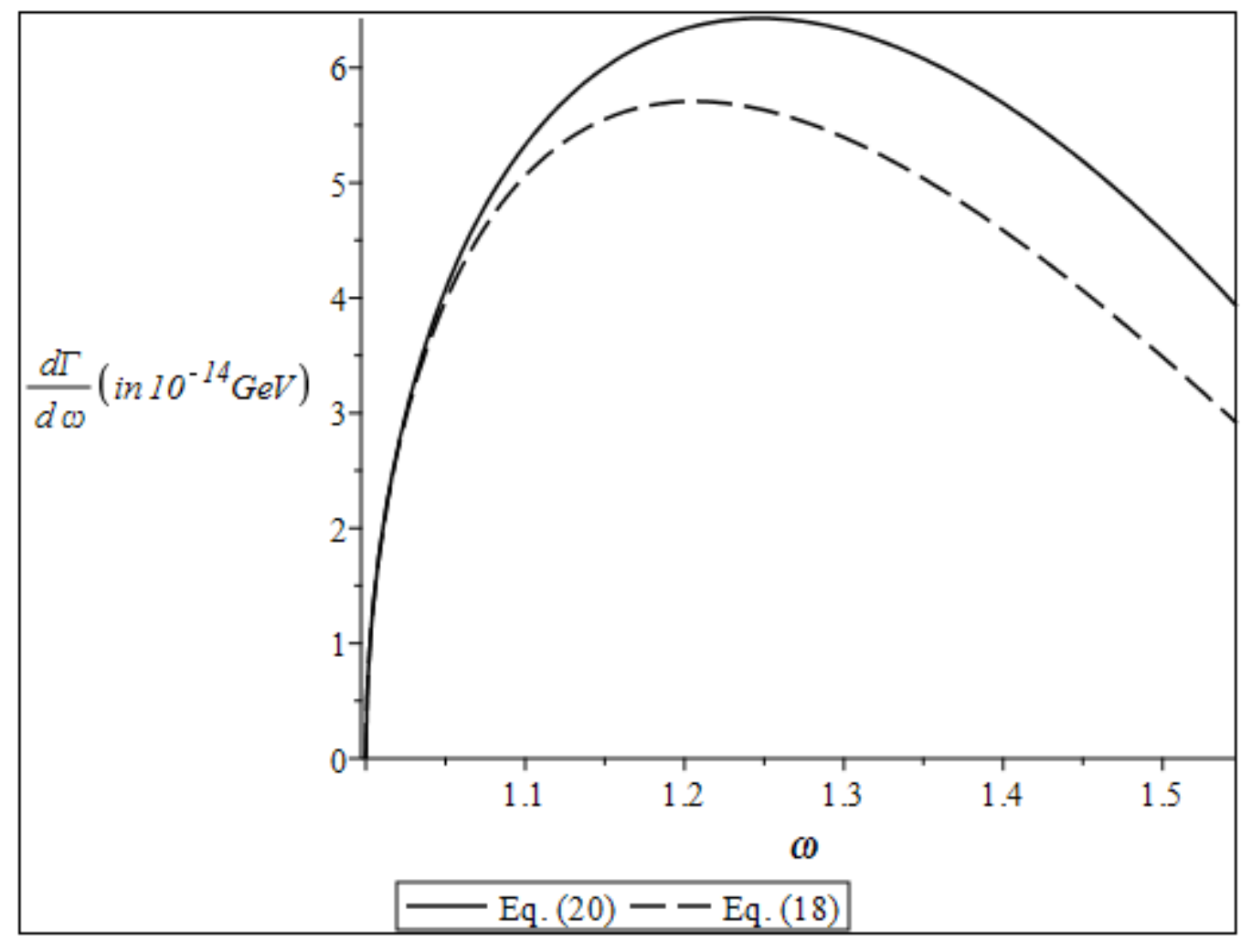}
\caption{Differential semileptonic decay widths of $B \to {D^*}\ell \nu $ versus $\omega $ for two cases of IWFs.}
\label{fig:BtoD*}
\end{figure}

\begin{figure}
\centering
\includegraphics[width=0.48\linewidth]{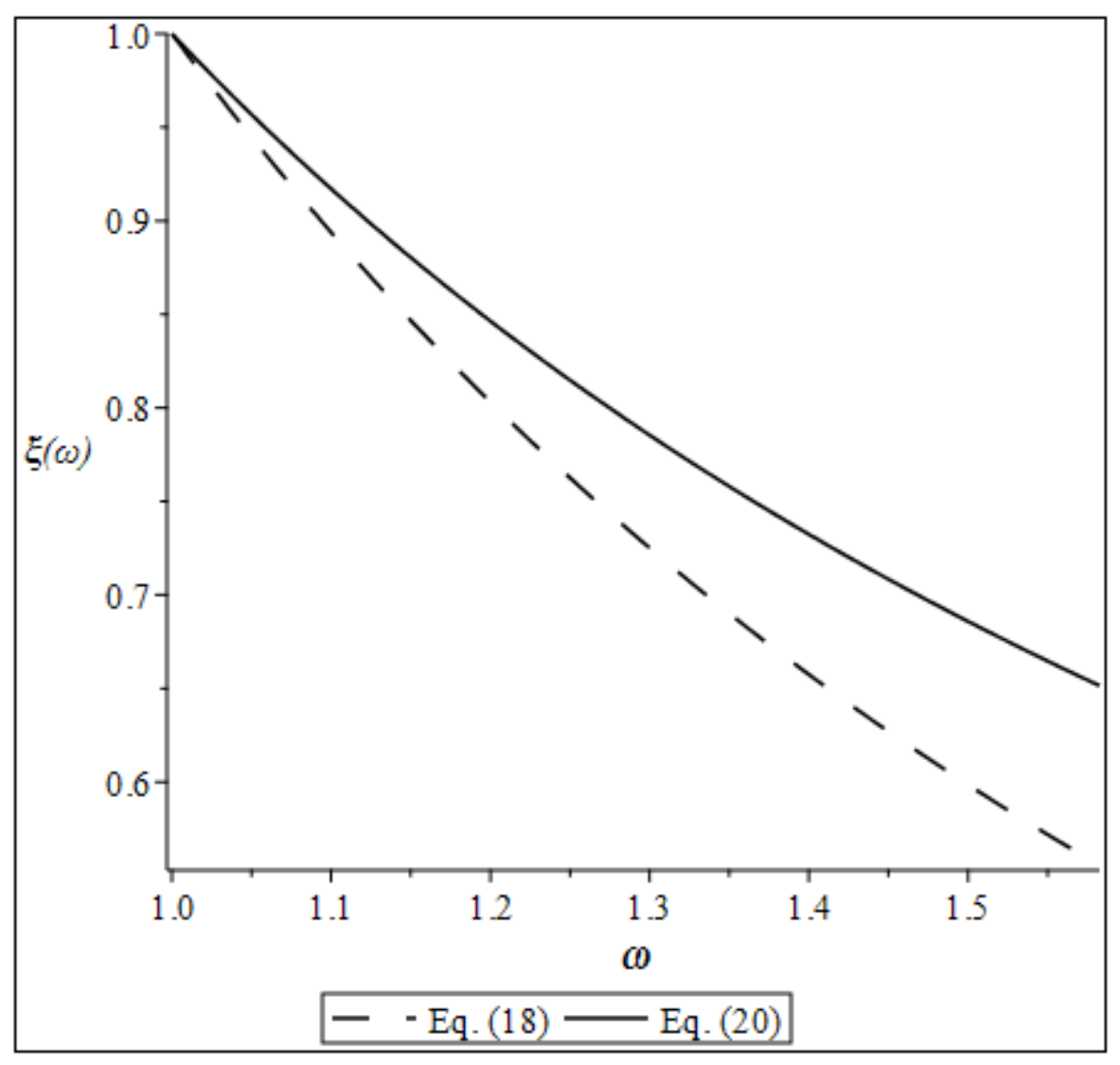}
\caption{Variation of IWFs.}
\label{fig:IWFs}
\end{figure}

IWF can be parameterized within a phenomenological model \cite{LeYaouanc:2003rn},
\begin{equation}
    \xi (\omega ) = {\left( {\frac{2}{{\omega  + 1}}} \right)^{2{\rho ^2}}}
    \label{eq:IWF18}
\end{equation}
where we take ${\rho ^2} = 1.15$ \cite{LeYaouanc:2003rn}. In the limit of zero lepton mass, we can write for the differential semileptonic decay width of ${\bar B_{(s)}} \to D_{(s)}^*\ell \bar \nu $ as \cite{Rahmani:2017vbg,Bernlochner:2012bc}
\begin{equation}
    \begin{gathered}
  \frac{{d\Gamma ({{\bar B}_{(s)}} \to D_{(s)}^*\ell \bar \nu )}}{{d\omega }} = \frac{{G_F^2}}{{48{\pi ^3}}}M_{D_{(s)}^*}^3{({M_{{B_{(s)}}}} - {M_{D_{(s)}^*}})^2}{[1 + {\beta ^{{A_1}}}(1)]^2} \times \sqrt {{\omega ^2} - 1} {(\omega  + 1)^2}|{V_{cb}}{|^2} \hfill \\
   \times \xi _{}^2(\omega )\left[ {1 + \frac{{4\omega }}{{\omega  + 1}}\frac{{M_{{B_{(s)}}}^2 - 2\omega {M_{{B_{(s)}}}}{M_{D_{(s)}^*}} + M_{D_{(s)}^*}^2}}{{{{({M_{{B_{(s)}}}} - {M_{D_{(s)}^*}})}^2}}}} \right]K(\omega ) \hfill 
   \label{eq:BstoDs*}
\end{gathered} 
\end{equation}
where ${\beta ^{{A_1}}}(1) =  - 0.01$  and $K(\omega )$=1. By integrating the differential semileptonic decay widths of Eqs. \eqref{eq:BtoD} and \eqref{eq:BstoDs*} over the range $1 \leqslant \omega  \leqslant \frac{{M_{{B_{(s)}}}^2 + M_{{D_{(s)}}}^2}}{{2{M_{{B_{(s)}}}}{M_{{D_{(s)}}}}}}$, using the life times of $B$ mesons, ${\tau _B} = 1.638ps,{\tau _{{B_s}}} = 1.512ps,$ and $|{V_{cb}}| = 0.04$, we can calculate the semileptonic branching fractions of $B$ mesons. If we consider IWF as
\begin{equation}
    \xi (\omega ) = \frac{{2(\omega  + 1)}}{{{{[\omega  + 1 + ({\rho ^2} - \frac{1}{2})(\omega  - 1)]}^2}}},
    \label{eq:IWF20}
\end{equation}
 with ${\rho ^2} = 0.9$ \cite{Coleman:2000gu}, we get $\Gamma (\bar B \to {D^*}l\bar \nu ) = {\text{2}}{\text{.934}} \times {10^{ - 14}}GeV$ and $Br(\bar B \to {D^*}l\bar \nu ) = 7.31\,\% $ . Also we have obtained $\Gamma (\bar B \to \bar Dl\bar \nu ) = {\text{1}}{\text{.019}} \times {10^{ - 14}}GeV$ and $Br(\bar B \to \bar Dl\bar \nu ) = 2.54\% $. Fig \ref{fig:BtoD*} shows the differential semileptonic decay widths of $B \to {D^*}\ell \nu $ for two cases of IWFs. As we can see from Fig. \ref{fig:IWFs}, IWFs are normalized to unity at $\omega  = 1$. Since one can see that $\frac{{d\Gamma }}{{d\omega }}$ for Eq. \eqref{eq:IWF20} grows faster than Eq. \eqref{eq:IWF18}, we can expect that the branching ratios for the case of \eqref{eq:IWF20} be larger than the case of Eq. \eqref{eq:IWF18}. 
 We have shown our results in Table \ref{tab:semi} using two IWFs with the comparison of the other results.
 
\begin{center}

\begin{table}[ht]
\caption{Semileptonic decay widths of $B$ mesons}
\label{tab:semi}
\begin{tabularx}{\textwidth}{|X|X|X|X|X|X|} 
\hline    

Decay & $\Gamma$ (in GeV) (with Eq. \eqref{eq:IWF20}) & $\Gamma$ (in GeV) (with Eq. \eqref{eq:IWF18}) & $Br$ (Ours) (with Eq.\eqref{eq:IWF20}) & $Br$ (Ours) (with Eq.\eqref{eq:IWF18}) & $Br$ (Others) \\  \hline

${\bar B_s} \to {\bar D_s}\ell \bar \nu $ & ${\text{1}}{\text{.089}} \times {10^{ - 14}}$ & ${\text{0}}{\text{.894}} \times {10^{ - 14}}$ & 2.50 & 2.05 & $2.1 \pm 0.2$ \cite{Faustov:2012mt}  \\ \hline

$\bar B \to \bar D\ell \bar \nu$ & ${\text{1}}{\text{.019}} \times {10^{ - 14}}$ & ${\text{0}}{\text{.821}} \times {10^{ - 14}}$ & 2.54 & 2.04 & $2.35 \pm 0.09$ \cite{PDG:2020}, ${\text{2}}{\text{.31}} \pm {\text{0}}{\text{.09}}$ \cite{Ivanov:1998ya}, $1.8 \pm 0.5$ \cite{Hioki:1993yr} \\  \hline

$\bar B \to {D^*}l\bar \nu $ & ${\text{2}}{\text{.934}} \times {10^{ - 14}}$ &  ${\text{2}}{\text{.535}} \times {10^{ - 14}}$   &   7.31   &   6.31   &   $5.66 \pm 0.22$ \cite{PDG:2020}, $4.9 \pm 0.8$ \cite{Hioki:1993yr} \\ \hline

${\bar B_s} \to D_s^*l\bar \nu $ & ${\text{3}}{\text{.127}} \times {10^{ - 14}}$ &    ${\text{2}}{\text{.734}} \times {10^{ - 14}}$   &   7.19   &   6.28   &  $5.3 \pm 0.5$ \cite{Faustov:2012mt}, $7.49 - 7.66$ \cite{Zhao:2006at} \\  \hline
 
\end{tabularx} 
\end{table}
\end{center}

From the results in Table \ref{tab:semi}, we have found the ratios of $B$ semileptonic decays as $\frac{{\Gamma (\bar B \to {D^*}l\bar \nu )}}{{\Gamma (\bar B \to D\ell \bar \nu )}} = {\text{2}}{\text{.88}}$ using of Eq. \eqref{eq:IWF20} for IWF, $\frac{{\Gamma (\bar B \to {D^*}l\bar \nu )}}{{\Gamma (\bar B \to D\ell \bar \nu )}} = {\text{3}}{\text{.09}}$ considering IWF as Eq. \eqref{eq:IWF18}, $\frac{{\Gamma ({{\bar B}_s} \to D_s^*l\bar \nu )}}{{\Gamma ({{\bar B}_s} \to {D_s}\ell \bar \nu )}} = {\text{2}}{\text{.87}}$ with Eq. \eqref{eq:IWF20} and $\frac{{\Gamma ({{\bar B}_s} \to D_s^*l\bar \nu )}}{{\Gamma ({{\bar B}_s} \to {D_s}\ell \bar \nu )}} = {\text{3}}{\text{.06}}$ with Eq. \eqref{eq:IWF18}, which are in accordance with the ones obtained in Ref. \cite{Bowler:1995bp}, $\frac{{\Gamma (\bar B \to {D^*}l\bar \nu )}}{{\Gamma (\bar B \to D\ell \bar \nu )}} = 3.2_{ - 2}^{ + 3} \pm 1.0$  and $\frac{{\Gamma ({{\bar B}_s} \to D_s^*l\bar \nu )}}{{\Gamma ({{\bar B}_s} \to {D_s}\ell \bar \nu )}} = {\text{3}}{\text{.3}}_{ - 1}^{ + 2} \pm 1.0$, respectively. In the case of $\frac{{\Gamma (\bar B \to {D^*}l\bar \nu )}}{{\Gamma (\bar B \to D\ell \bar \nu )}}$, the differences of our values with them are about 10.05 $\%$ with Eq. \eqref{eq:IWF20} and 3.56 $\%$ with Eq. \eqref{eq:IWF18}, respectively. For the ratios $\frac{{\Gamma ({{\bar B}_s} \to D_s^*l\bar \nu )}}{{\Gamma ({{\bar B}_s} \to {D_s}\ell \bar \nu )}}$, we have obtained the differences about 12.99 $\%$ with Eq. \eqref{eq:IWF20} and 7.31 $\%$ with Eq. \eqref{eq:IWF18} in comparison with Ref. \cite{Bowler:1995bp}. 

Moreover, our branching ratios of semileptonic $B$ decays are close to the reported values of Hiller et. al., which are $Br({B^0} \to {D^ + }(e,\mu )\nu ) = (2.23 \pm 0.24) \times {10^{ - 2}}$ and $Br({B^0} \to {D^{ + *}}(e,\mu )\nu ) = (5.34 \pm 0.40) \times {10^{ - 2}}$ \cite{Hiller:2016kry}. The deviations of our values with them are $\pm0.138$, $\pm0.083$ with Eqs. \eqref{eq:IWF20} and \eqref{eq:IWF18}, respectively, for $B$ to $D$ semileptonic decay, and $\pm0.368$ with Eq. \eqref{eq:IWF20}, $\pm0.182$ with Eq. \eqref{eq:IWF18} for $B$ to $D^*$ semileptonic decay.

In the heavy quark limit, Ivanov et. al. obtained the semileptonic decay branching fractions of $B$ mesons 2.65 for ${B^0} \to {D^ + }{\ell ^ - }\bar \nu $  and 7.21 for ${B^0} \to {D^{* + }}{\ell ^ - }\bar \nu $ \cite{Ivanov:2015tru}. Our results of $Br({B^0} \to {D^ + }{\ell ^ - }\bar \nu ) = 2.54$ and $Br({B^0} \to {D^{* + }}{\ell ^ - }\bar \nu ) = 7.31$ are in a good agreement with them having about 4.23 $\%$ and 1.32 $\%$ differences, respectively, as well as with Ref. \cite{Azizi:2008vt} in which they reported: $Br({B^0} \to {D^{* + }}{\ell ^ - }\bar \nu ) = 4.57 - 9.12$. In the case of branching ratios $B_s$, Azizi reported $Br({B_s} \to {D_s}\ell \nu ) = 2.8 - 3.5$ \cite{Azizi:2008tt} and $Br({B_s} \to D_s^*\ell \nu ) = 1.89 - 6.61$ \cite{Azizi:2008vt} in the framework of three point QCD sum rules. Our results are competible with theirs well, where the deviations of our results are about $\pm0.106$ and $\pm0.049$ for the branching ratios of ${B_s} \to {D_s}\ell \nu $ and ${B_s} \to D_s^*\ell \nu $, respectively, comparing with Refs. \cite{Azizi:2008tt,Azizi:2008vt}.

\section{Nonleptonic decays of $B$ and $D$ mesons}
\begin{figure}
\centering
\includegraphics[width=0.48\linewidth]{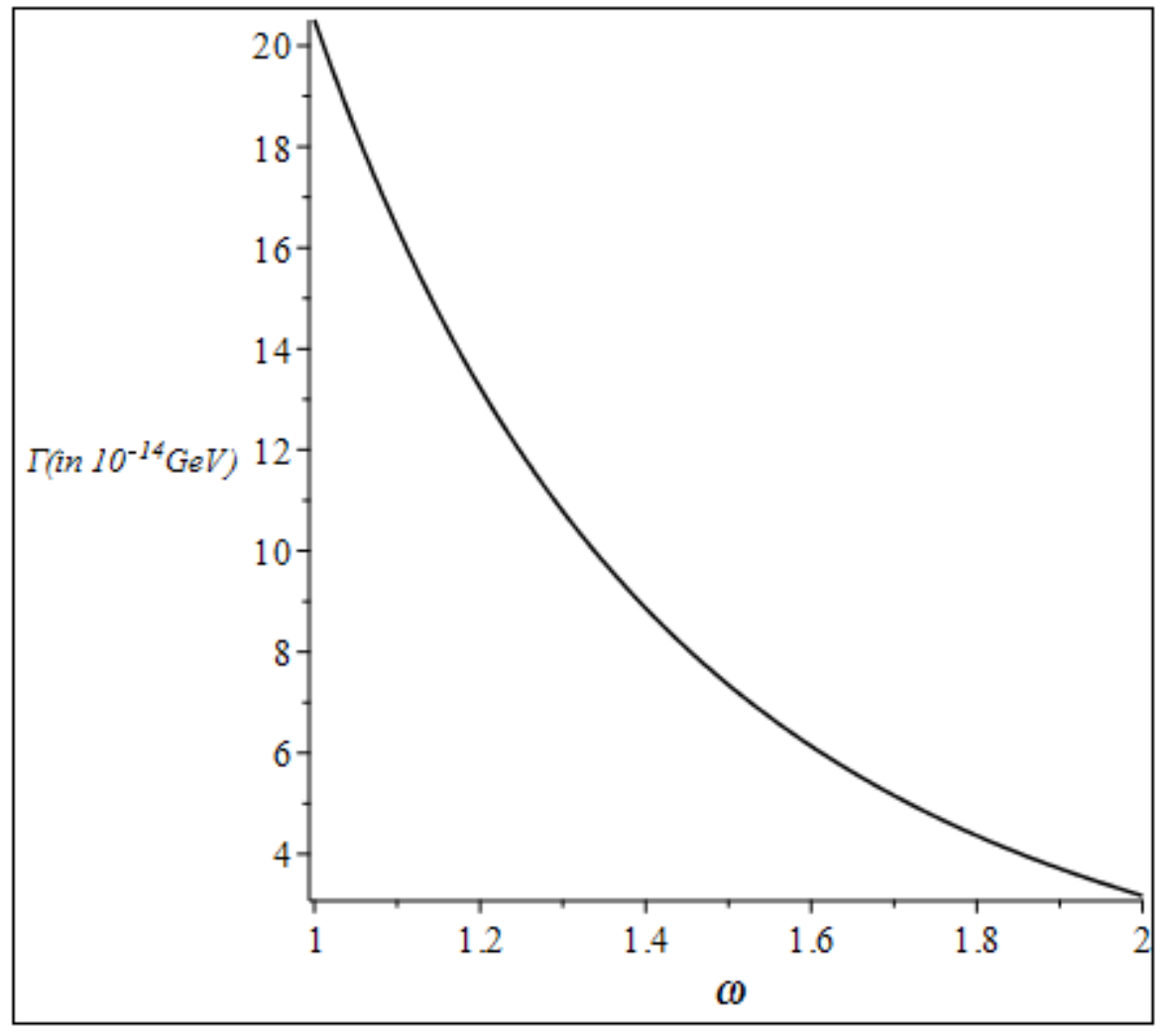}
\caption{Decay width of $D \to K\pi $ versus $\omega$ (using Eq. \eqref{eq:IWF18} for IWF).}
\label{fig:DtoKp}
\end{figure}

The flavor changing decays of the $b$-meson and $c$-meson can be used as a test of SM. Here we want to study two-body nonleptonic decays of $B$ and $D$ mesons through to a hadronic state and a pion meson in their final processes. In each channel the nonleptonic decay is related to the semileptonic differential decay at maximal recoil as follows \cite{Bernlochner:2012bc}
\begin{equation}
    \Gamma (B \to {D^{(*)}}\pi ) = \frac{{3{\pi ^2}{C^2}|{V_{ud}}{|^2}f_\pi ^2}}{{{M_B}{M_{{D^{(*)}}}}}}\frac{{d\Gamma (B \to {D^{(*)}}\ell \bar \nu )}}{{d\omega }}{|_{{\omega _{\max }}}},
    \label{eq:BtoDp}
\end{equation}
where $C|{V_{ud}}| \approx 1$  and the dot product $\omega  = v.v'$  is determined by considering momentum conservation of two-body decays 
\begin{equation}
    \omega  = \frac{{M_B^2 + M_{{D^{(*)}}}^2 - M_\pi ^2}}{{2{M_B}{M_{{D^{(*)}}}}}}.
\end{equation}
The decay widths of $D$ are given by \cite{Shah:2014yma},
\begin{equation}
    \Gamma ({D^0} \to {K^ - }{\pi ^ + }) = {C_f}\frac{{G_F^2|{V_{cs}}{|^2}|{V_{ud}}{|^2}f_\pi ^2}}{{32\pi M_{{D_s}}^3}}{\left( {\lambda (M_D^2,M_{{K^ - }}^2,M_\pi ^2)} \right)^{\frac{3}{2}}}f_ + ^2({q^2}),
    \label{eq:DtoK-p+}
\end{equation}
for the case of $c \to s + u + \bar d$  and
\begin{equation}
    \Gamma ({D^0} \to {K^ + }{\pi ^ - }) = {C_f}\frac{{G_F^2|{V_{cd}}{|^2}|{V_{us}}{|^2}f_\pi ^2}}{{32\pi M_{{D_s}}^3}}{\left( {\lambda (M_D^2,M_{{K^ + }}^2,M_\pi ^2)} \right)^{\frac{3}{2}}}f_ + ^2({q^2}),
    \label{eq:DtoK+p-}
\end{equation}
for the case of $c \to d + u + \bar d$, where the factor $\lambda (M_D^2,M_{{K^ + }}^2,M_\pi ^2)$ is the usual K\"allen triangle function and given by
\begin{equation}
    \lambda (x,y,z) = {x^2} + {y^2} + {z^2} - xy - yz - zx,
\end{equation}
and the color factor is ${C_f} = C_A^2 + C_B^2$,  where
\begin{align}
{C_A} &= \frac{1}{2}({C_ + } + {C_ - }),{C_B} = \frac{1}{2}({C_ + } - {C_ - }),  \\
    {C_ + } &= 1 - \frac{{{\alpha _s}}}{\pi }\log \left( {\frac{{{m_W}}}{{{m_c}}}} \right), {C_ - } = 1 + 2\frac{{{\alpha _s}}}{\pi }\log \left( {\frac{{{m_W}}}{{{m_c}}}} \right).
\end{align}
Further we get \cite{Ivanov:2015tru,Ivanov:2016qtw}
\begin{equation}
    {f_ + }({q^2}) = \xi (\omega )\frac{{{M_D} + {M_\phi }}}{{2\sqrt {{M_D}{M_\phi }} }}
\end{equation}
for the weak transition form factor which is related to IWF. In Fig. \ref{fig:DtoKp}, we show the decay width of $D \to K\pi $  as a function of $\omega$. Considering two cases for IWF in Eqs \eqref{eq:IWF18} and \eqref{eq:IWF20}, taking ${\tau _{{D^0}}} = 0.410p{s^{ - 1}}$, ${M_\phi } = 1.019GeV,\,{M_\pi } = 0.139GeV$, ${M_K} = 0.{\text{493}}GeV, {f_\pi } = 0.130GeV$, $|{V_{us}}| = 0.{\text{225}}$ and $|{V_{ud}}| = 0.{\text{974}}$, we obtain the nonleptonic decay rates with Eqs. \eqref{eq:DtoK-p+}, \eqref{eq:DtoK+p-} as well as Eq. \eqref{eq:BtoDp} and show them in Table \ref{tab:nonleptonic}, where one can see that our results are consistent with the other results in Refs. \cite{Fusheng:2011tw,Kramer:1997yh,PDG:2020}.

\begin{center}

\begin{table}[ht]
\caption{Nonleptonic decay rates of $B$ and $D$ mesons}
\label{tab:nonleptonic}
\begin{tabularx}{\textwidth}{|X|X|X|X|X|} 
\hline    

Decay & $Br$ (Ours) (with Eq.\eqref{eq:IWF20}) & $Br$ (Ours) (with Eq.\eqref{eq:IWF18}) & $Br$ (Exp. \cite{PDG:2020}) & $Br$ (Others) \\  \hline

$B \to {D^*}\pi $ & ${\text{4}}{\text{.811}} \times {10^{ - 3}}$ & ${\text{3}}{\text{.571}} \times {10^{ - 3}}$ & $(4.90 \pm 0.17) \times {10^{ - 3}}$ & $4.74 \times {10^{ - 3}}$ \cite{Kramer:1997yh} \\ \hline

$B \to D\pi $ & ${\text{4}}{\text{.867}} \times {10^{ - 3}}$ & ${\text{3}}{\text{.539}} \times {10^{ - 3}}$ & $(4.68 \pm 0.13) \times {10^{ - 3}}$ & $5.91 \times {10^{ - 3}}$ \cite{Kramer:1997yh} \\  \hline

${D^0} \to {K^ - }{\pi ^ + }$ & $3.370\%$  &   $1.914\% $ & $(3.950 \pm 0.031)\% $   &   $4.03\% $ \cite{Fusheng:2011tw} \\  \hline
 
 ${D^0} \to {K^ + }{\pi ^ - }$ & $0.965 \times {10^{ - 4}}$ & ${\text{0}}{\text{.548}} \times {10^{ - 4}}$ & - & $(1.12 \pm 0.05) \times {10^{ - 4}}$ \cite{Fusheng:2011tw} \\  \hline
\end{tabularx} 
\end{table}
\end{center}
\par
Besides, our results of $Br({B^0} \to {D^{* + }}{\pi ^ - }) = 0.357\% $ and $Br({B^0} \to {D^ + }{\pi ^ - }) = 0.354\% $ are in agreement with the ones obtained in Ref. \cite{AbdElHady:1997rj}, $Br({B^0} \to {D^ + }{\pi ^ - }) = 0.345\% $ and $Br({B^0} \to {D^{* + }}{\pi ^ - }) = 0.331\% $,  with differences about 7.85 $\%$ and 2.61 $\%$, respectively. In the case of charm nonleptonic decay ${D^0} \to {K^ - }{\pi ^ + }$,  we have obtained the deviation as $\pm0.147$ compared to PDG \cite{PDG:2020}.

\section{Conclusion}
In the present work, we have presented a phenomenological model based on Screened potential and obtained the energy, the wave functions and the masses for the beauty and charm mesons using a variational method. Consequently, we study the decay properties of the  heavy-light mesons, such as the leptonic decay, the radiative leptonic decay, the semileptonic decay and two-body nonleptonic decay. Two forms of IWFs were considered. Through this work, we have gotten the results of Tables \ref{tab:mass}--\ref{tab:nonleptonic}, where the results we provided give a satisfactory description of properties of beauty and charm mesons and compatible with the other theoretical or experimental results. Thus, our results can be useful for further studying of the properties of $B$ and $D$ mesons and their branching fractions for the leptonic decay, the radiative decay, the semileptonic decay and the nonleptonic decay.


\begin{thebibliography}{10}


\bibitem{Fusheng:2011tw} 
  F. S. Yu, X.~X.~Wang and C.~D.~L\"u,
  Phys.\ Rev.\ D {\bf 84}, 074019 (2011)
  [arXiv:1101.4714 [hep-ph]].



\bibitem{Hioki:1993yr} 
  Z. Hioki, 
  Z. \ Phys.\ C {\bf 62}, 633-637 (1994)
  
  

\bibitem{LeYaouanc:2003rn}
A.~Le Yaouanc, L.~Oliver and J.~Raynal,
Phys. Rev. D \textbf{69}, 094022 (2004)
[arXiv:hep-ph/0307197 [hep-ph]].



\bibitem{Coleman:2000gu}
T.~Coleman, M.~G.~Olsson and S.~Veseli,
Phys. Rev. D \textbf{63}, 032006 (2001)
[arXiv:hep-ph/0009103 [hep-ph]].



\bibitem{Zhao:2006at}
S.~M.~Zhao, X.~Liu and S.~J.~Li,
Eur. Phys. J. C \textbf{51}, 601-606 (2007)
[arXiv:hep-ph/0612008 [hep-ph]].



\bibitem{Paver:1992cb}
N.~Paver,
Nucl. Phys. B Proc. Suppl. \textbf{27}, 39-46 (1992)



\bibitem{Wang:2005qx}
G.~L.~Wang,
Phys. Lett. B \textbf{633}, 492-496 (2006)
[arXiv:math-ph/0512009 [math-ph]].



\bibitem{Yang:2011ie}
M.~Z.~Yang,
Eur. Phys. J. C \textbf{72}, 1880 (2012)
[arXiv:1104.3819 [hep-ph]].



\bibitem{Chang:2018aut}
Q.~Chang, X.~N.~Li, X.~Q.~Li and F.~Su,
Chin. Phys. C \textbf{42}, no.7, 073102 (2018)
[arXiv:1805.00718 [hep-ph]].



\bibitem{Azizi:2008tt}
K.~Azizi,
Nucl. Phys. B \textbf{801}, 70-80 (2008)
[arXiv:0805.2802 [hep-ph]].



\bibitem{Azizi:2008vt}
K.~Azizi and M.~Bayar,
Phys. Rev. D \textbf{78}, 054011 (2008)
[arXiv:0806.0578 [hep-ph]].



\bibitem{Ivanov:1998ya}
M.~A.~Ivanov, J.~G.~K\"orner, V.~E.~Lyubovitskij and A.~Rusetsky,
Phys. Rev. D \textbf{59}, 074016 (1999)
[arXiv:hep-ph/9809254 [hep-ph]].



\bibitem{AbdElHady:1997rj}
A.~Abd El-Hady, A.~Datta and J.~Vary,
Phys. Rev. D \textbf{58}, 014007 (1998)
[arXiv:hep-ph/9711338 [hep-ph]].



\bibitem{Lu:2002mn}
C.~D.~L\"u and G.~L.~Song,
Phys. Lett. B \textbf{562}, 75-80 (2003)
[arXiv:hep-ph/0212363 [hep-ph]].



\bibitem{Li:2009wq}
R.~H.~Li, C.~D.~L\"u and Y.~M.~Wang,
Phys. Rev. D \textbf{80}, 014005 (2009)
[arXiv:0905.3259 [hep-ph]].



\bibitem{Ivanov:2015tru}
M.~A.~Ivanov, J.~G.~K\"orner and C.~T.~Tran,
Phys. Rev. D \textbf{92}, no.11, 114022 (2015)
[arXiv:1508.02678 [hep-ph]].



\bibitem{Kramer:1997yh}
G.~Kramer and C.~D.~L\"u,
Int. J. Mod. Phys. A \textbf{13}, 3361-3384 (1998)
[arXiv:hep-ph/9707304 [hep-ph]].



\bibitem{Sun:2019xyw}
H.~K.~Sun and M.~Z.~Yang,
Phys. Rev. D \textbf{99}, no.9, 093002 (2019)
[arXiv:1903.04295 [hep-ph]].


\bibitem{Devlani:2012zz}
N.~Devlani and A.~Rai,
Eur. Phys. J. A \textbf{48}, 104 (2012)



\bibitem{Pang:2019ttv}
C.~Q.~Pang,
Phys. Rev. D \textbf{99}, no.7, 074015 (2019)
[arXiv:1902.02206 [hep-ph]].



\bibitem{Chang:2014jca}
C.~Chang, H.~F.~Fu, G.~L.~Wang and J.~M.~Zhang,
Sci. China Phys. Mech. Astron. \textbf{58}, no.7, 071001 (2015)
[arXiv:1411.3428 [hep-ph]].



\bibitem{Pathak:2012kp}
K.~K.~Pathak, D.~Choudhury and N.~Bordoloi,
Int. J. Mod. Phys. A \textbf{28}, 1350010 (2013)
[arXiv:1207.1849 [hep-ph]].



\bibitem{Chakrabarty:1990gs}
S.~Chakrabarty and S.~Deoghuria,
J. Phys. G \textbf{16}, 185-193 (1990)



\bibitem{Hassanabadi:2016kqq}
H.~Hassanabadi, M.~Ghafourian and S.~Rahmani,
Few Body Syst. \textbf{57}, no.4, 249-254 (2016)



\bibitem{PDG:2020}
P. A. Zyla et al., (Particle Data Group),
to be published in Prog. Theor. Exp. Phys. \textbf{2020}, 083C01 (2020)



\bibitem{Rahmani:2017vbg}
S.~Rahmani and H.~Hassanabadi,
Eur. Phys. J. A \textbf{53}, no.9, 187 (2017)



\bibitem{Ebert:2006hj}
D.~Ebert, R.~N.~Faustov and V.~O.~Galkin,
Phys. Lett. B \textbf{635}, 93-99 (2006)
[arXiv:hep-ph/0602110 [hep-ph]].



\bibitem{Lubicz:2016bbi}
V.~Lubicz, A.~Melis and S.~Simula,
PoS \textbf{LATTICE2016}, 291 (2017)
[arXiv:1610.09671 [hep-lat]].



\bibitem{Fleischer:2019wlx}
R.~Fleischer, R.~Jaarsma and G.~Koole,
Eur. Phys. J. C \textbf{80}, no.2, 153 (2020)
[arXiv:1912.08641 [hep-ph]].



\bibitem{Buras:2003td}
A.~J.~Buras,
Phys. Lett. B \textbf{566}, 115-119 (2003)
[arXiv:hep-ph/0303060 [hep-ph]].



\bibitem{Hulsbergen:2013lma}
W.~Hulsbergen,
Mod. Phys. Lett. A \textbf{28}, 1330023 (2013)
[arXiv:1306.6474 [hep-ph]].



\bibitem{Pathak:2011km}
K.~K.~Pathak and D.~Choudhury,
Chin. Phys. Lett. \textbf{28}, 101201 (2011)
[arXiv:1108.5315 [hep-ph]].



\bibitem{Rahmani:2017exi}
S.~Rahmani and H.~Hassanabadi,
Few Body Syst. \textbf{58}, no.5, 150 (2017)



\bibitem{Hassanabadi:2014isa}
H.~Hassanabadi, S.~Rahmani and S.~Zarrinkamar,
Eur. Phys. J. C \textbf{74}, no.10, 3104 (2014)
[arXiv:1407.3901 [hep-ph]].



\bibitem{Bernlochner:2012bc}
F.~U.~Bernlochner, Z.~Ligeti and S.~Turczyk,
Phys. Rev. D \textbf{85}, 094033 (2012)
[arXiv:1202.1834 [hep-ph]].



\bibitem{Faustov:2012mt}
R.~N.~Faustov and V.~O.~Galkin,
Phys. Rev. D \textbf{87}, no.3, 034033 (2013)
[arXiv:1212.3167 [hep-ph]].



\bibitem{Bowler:1995bp}
K.~C.~Bowler \textit{et al.} [UKQCD],
Phys. Rev. D \textbf{52}, 5067-5094 (1995)
[arXiv:hep-ph/9504231 [hep-ph]].



\bibitem{Hiller:2016kry}
G.~Hiller, D.~Loose and K.~Sch\"onwald,
JHEP \textbf{12}, 027 (2016)
[arXiv:1609.08895 [hep-ph]].



\bibitem{Shah:2014yma}
M.~Shah, B.~Patel and P.~Vinodkumar,
Eur. Phys. J. C \textbf{76}, no.1, 36 (2016)
[arXiv:1412.7400 [hep-ph]].



\bibitem{Ivanov:2016qtw}
M.~A.~Ivanov, J.~G.~K\"orner and C.~T.~Tran,
Phys. Rev. D \textbf{94}, no.9, 094028 (2016)
[arXiv:1607.02932 [hep-ph]].



\end{thebibliography}
\end{document}